# Ultralight High-Entropy Nanowire Scaffolds for Extreme-Temperature Functionality

Cameron S. Jorgensen,[1,2,3] Corisa Kons,[1] William Stallions,[1] Austin C. Houston,[1] Gerd Duscher,[1] Dustin A. Gilbert[1] [†]

[1] Materials Science and Engineering, University of Tennessee, Knoxville, TN, 37996, United States
[2] Air Force Research Laboratory, Wright Patterson Air Force Base, OH, 45433, United States
[3] Arctos Technology Solutions, Dayton OH, 45431, United States

[†] Email: dagilbert@utk.edu

**Abstract**
High-entropy alloys (HEAs) combine compositional disorder with exceptional functional tunability, yet their inherently high-density limits use in lightweight systems. Here, we introduce entropy-architected nanowire metamaterials, a class of materials that couple configurational entropy with structural porosity to achieve metal-like functionality at ultralow density. FeCoNiCrCu HEA nanowires were electrodeposited into porous templates and freeze-cast into three-dimensional "bird's-nest" scaffolds with densities below 1 % of the bulk metal. The resulting architectures retain a disordered face-centered-cubic phase, exhibit Curie temperatures exceeding 1000 K, and deliver thermal diffusivity ($\approx$ 0.211 mm$^2$ s$^{-1}$) comparable to titanium alloys. Structural and spectroscopic analyses reveal nanoscale Cu segregation that enhances magnetic ordering and thermal stability. These findings demonstrate that configurational entropy and architectural hierarchy can be co-engineered to yield lightweight, high-temperature functional materials for extreme-environment applications.

## Introduction

High entropy alloys (HEAs) are a rapidly expanding class of materials that exploit configurational entropy to stabilize complex solid solutions.[1-3] This enables the combination of elements, including immiscible binary pairs, to form a single-phase structure with highly disordered atomic environments. The resulting distribution of atomic radii and bond lengths generates localized lattice distortions – so-called "extremities" – which drive a range of functional behaviors, including tunable magnetism, electrical transport, and thermal properties.[3-8] In addition to these emergent functionalities, HEAs are valued for their mechanical robustness, high-temperature stability, corrosion resistance, and – in many cases – their affordability and recyclability.[9]

These combined attributes make HEAs appealing for both structural and functional technologies, from hard and soft magnetic materials to catalysts and aerospace components.[10-12] However, their widespread adoption in weight-sensitive systems remains constrained by a key limitation: density. Most HEAs have densities above 8 g/cm$^3$, roughly twice that of titanium and triple that of aluminum, which severely restricts their use in weight-critical applications.

Nanostructured metamaterials offer a promising route to overcome this limitation. Architectures such as sol-gel or freeze-cast nanowire scaffolds can dramatically reduce the overall density, with values tunable from bulk-like down to <10 mg/cm$^3$, while preserving essential mechanical properties. These materials – like carbon, silica, and alumina aerogels – are defined by their high surface-area-to-volume (SA:V) ratios ($\approx$1.5$\times$10$^6$ m$^{-1}$) and ultralow densities,[13] and have been widely investigated for use in electronics [1,2], novel computing,[14, 15] catalysis,[16, 17] energy storage,[18-21] memristors[15], unique magnetism[22], lightweight structural systems,[20, 23] filters,[24] electromagnetic shielding,[25] as well as hydrogenation materials.[13]

HEAs are especially well-suited to this metamaterial strategy. Many HEAs are composed of 3d transition metals, which are readily electrodeposited into nanowires. This opens a scalable and controllable route to fabricate HEA-based scaffolds – something not easily achieved with aluminum or titanium, which resist electrochemical processing. The resulting structures combine the strength and tunability of HEAs with the mass efficiency of porous architectures. Moreover, nanowire scaffolds are known to retain a significant fraction of their intrinsic mechanical strength despite drastic reductions in density, further enhancing their suitability for demanding environments.

By integrating the compositional versatility of HEAs with the geometric control of metamaterials, it becomes possible to engineer low-density, high-performance materials with customizable functionality – addressing long-standing challenges in aerospace, catalysis, and thermal management.

In this work, we prepare nanowires of the high entropy alloy FeCoNiCrCu – a particularly promising composition known for its strength, corrosion resistance, and well-studied bulk behavior – and assemble them into ultra-low-density nanowire scaffolds. These scaffolds are then sintered to weld the wires together, enhancing their mechanical integrity without sacrificing porosity. The as-prepared and heat-treated structures are characterized using transmission electron microscopy (TEM) and energy loss spectroscopy (EELS), including in situ evaluation during high-temperature processing. Functional performance is assessed through high-temperature magnetometry and thermal conductivity measurements. Together, these results demonstrate a path to merging the compositional complexity of HEAs with the architectural tunability of metamaterials, enabling a new class of mechanically robust, lightweight materials for extreme-environment applications.

## **Methods**

High entropy FeCoNiCrCu nanowires were fabricated by electrodeposition from an aqueous solution of 0.08 M $CuSO_4$ + 0.16 M $CoSO_4$ + 0.16 M $FeSO_4$ + 0.21 M $NiSO_4$ + 0.08 M $CrCl_3$ + 0.8 M $BH_3O_3$ (1 M = 1 mol $L^{-1}$). The solution was mixed fresh before each growth. This formulation is closely related to the canonical Cantor alloy, Electrodeposition was performed at −2.0 V relative to a $Ag^+$/AgCl reference electrode into porous anodized aluminum oxide (AAO) or track-etched polycarbonate (PC) membranes[26, 27] in-which the back side was closed with a sputtered Au film; this film acts as the working electrode and the deposition occurs inside the pores of the membrane. Nanowires with diameters of 10–200 nm and lengths of 3–20 μm were achieved. The total deposited charge was 50 coulombs for 50 mm diameter membranes with 100 nm diameter pores, 60 μm in length. This value was determined by monitoring the instantaneous current for a growing increase in magnitude, indicating over-deposition. After deposition, the Au working electrode was selectively etched using a solution of 1:4:40 of $I_2$, KI, and $H_2O$. The AAO (PC) membranes were then dissolved by sonicating them in 6 M NaOH (chloroform). The nanowires were transferred to deionized water by using centrifugal precipitation to separate the nanowires from the solvent, then using a falcon pipette to remove the majority of the chloroform and resuspending the wires in methanol. Again using centrifugal precipitation, the wires were separated from the methanol, which was then replaced with de-ionized water. The water was replaced twice more, suspending the nanowires in nearly-pure de-ionized water. Nanowires were then freeze-cast into the scaffolded metamaterial. In this process, the nanowires dispersed in deionized water then allowed to settle or were precipitated by centrifugation. Next, excess water was decanted, leaving a volume of water and wires that was approximately the size of the final target scaffold. Then, the wires were re-dispersed by sonication to form a nanowire slurry, and quickly flash frozen using liquid nitrogen. This process promotes a uniform distribution of the wires with random orientations, forming a 'birds nest' architecture. The flash frozen structure was then placed in a freeze dryer and the ice was sublimated away, leaving the nanowire scaffold as a free standing structure. The solvent transfer and freeze-casting processes were described previously.[13, 28, 29] The mechanical strength of the nanowire scaffold can be further enhanced by sintering atmosphere at 600 °C, with an alternating atmosphere of air and forming gas.

Structural investigations were performed using X-ray diffraction, scanning electron microscopy (SEM), and high-resolution transmission electron microscopy (TEM) and were performed on both the scaffolds and individual wires. Select area electron diffraction (SAED) and electron energy loss spectroscopy (EELS) were performed during the TEM measurements to determine the local structure and compositional distribution, respectively; composition was also measured using energy dispersive X-ray spectroscopy. Thermal diffusivity measurements were performed at room temperature using a lock-in laser thermographer (1W, $\lambda$=455 nm). This technique heats the sample on one face using a laser with sinusoidal intensity, while the temperature on the opposite surface is monitored using a FLIR camera. By determining the phase delay between the laser power and the temperature on the opposite surface, the thermal diffusivity can be extracted,[30] and from that the thermal conductivity. Magnetometry measurements were performed between room temperature and 1000 K using a vibrating sample magnetometer (VSM).

## Results and Discussion
*Structure and Composition*

Polycrystalline FeCoNiCrCu nanowires were fabricated following the described recipe. Notably, this recipe did not work for the deposition of HEA thin films, these films were very Cr poor and Cu rich, but was effective for deposition into porous membranes to make nanowires. Reflecting on the described solution, Cr typically requires reducing agents to successfully achieve electrodeposition – agents which are absent in the current solution. In our solution, in nanowires, we observed that increasing the $CrCl_3$ also increased the Cu content in the resulting films and nanowires. Investigating the underlying chemistry, $2Cr + 3CuSO_4 \rightarrow Cr_2(SO_4)_3 + 3Cu$, $\Delta H$=-683 kJ/mol, an appreciably exothermic reaction. Thus, it becomes possible that the copper sulfate is spontaneously etching the Cr soon after deposition in the films, and is protected against this etching in the pores. One possibility is a strong dependence on the deposition current density – and deposition rate – effectively burying the Cr before it is etched. Alternatively, the limited mass transport in the pores reduces the ability of the $CuSO_4$ to access the deposited Cr. Cr plays a critical role in the high-temperature oxidation resistance of these materials, so achieving appreciable composition without complexing agents is an exciting development.

A TEM micrograph of the as-grown, electrodeposited and harvested wires is shown in Fig. 1A. The EELS images in Fig. 1B-F show the compositional distribution of each of the species are well distributed along the wire, with no apparent phase separation. The composition of the sample was measured with EDX to be $Fe_{18}Co_{18}Ni_{27}Cr_{25}Cu_{12}$. X-ray diffraction images (not shown) of the as-prepared wires indicated the HEA has a face centered cubic (FCC) structure, in agreement with the bulk material.[12] Local area diffraction shown in Fig. 1G and H confirm the structure and further indicate that the wires are polycrystalline with no preferential growth direction. Analyzing the (200) diffraction peak which occurs at 4.17 nm$^{-1}$ with a full width at half maximum of 1.42 nm$^{-1}$, corresponding to an interplanar spacing of 2.4 ± 0.7 Å, and a lattice parameter of 4.8 Å.

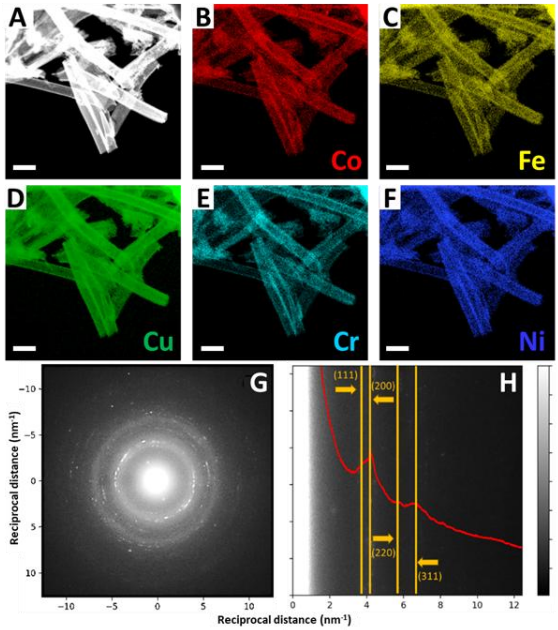

**Figure 1** TEM images of as-grown FeCoNiCrCu nanowires, showing (A) high resolution structure, and (B-F) compositional maps of Co, Fe, Cu, Cr and Ni, respectively, as measured with EELS. Selected area electron diffraction (SAED) for the nanowires (G) 2D diffraction pattern and (H) radial average profile.

The nanowires were annealed at 1000 °C to achieve crosslinking and improve structural integrity; the yield strength of sintered nanowires scaffolds made of Cu was previously reported to be between 4-100 kPa for 1-5 wt.%.[24] After annealing the nanowires at 1000 °C, the TEM images, Fig. 2A, and EELS images, Fig. 2B-F, were again captured, and notable phase separation is observed. In-particular, in the Cu EELS micrograph, Fig. 2D, the annealing process causes Cu to precipitate from the HEA and form particles on the surface of the sample. Previous works have also reported the precipitation of Cu in HEAs, especially in high temperature environments.[31] However, even with this precipitate, some of the Cu remains in the HEA lattice, again as observed in the EELS image. Using high resolution TEM to investigate the morphology of the surface further clarifies that the as-grown nanowires are smooth, Fig. 2G, but after annealing the surface is decorated with ≈15 nm diameter grains, Fig. 2H; these grains are likely Cu precipitates.

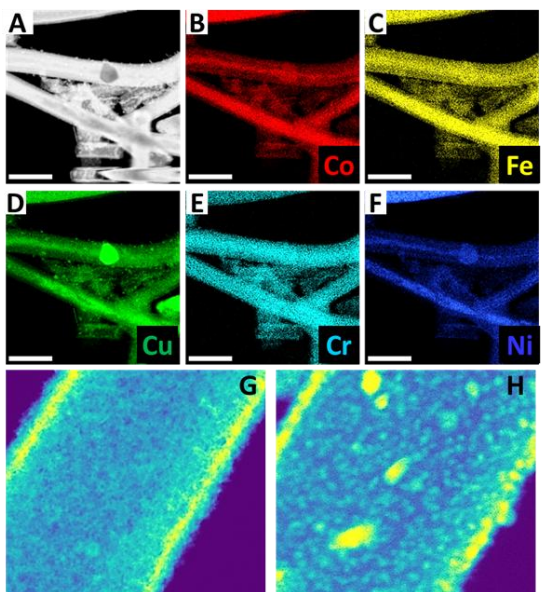

**Figure 2** TEM images of annealed FeCoNiCrCu nanowires, showing (A) high resolution structure, and (B-F) compositional maps of Co, Fe, Cu, Cr and Ni, respectively, as measured with EELS. The scale bar indicates 1 μm. TEM microscopy image of (G) an as-deposited nanowire with 10 nm polycrystalline microstructure and the (H) annealed nanowires, emphasizing the 15 nm Cu precipitates.

Nanowires of FeCoNiCrCu were prepared with a diameter of 250 nm and lengths of 10 μm and used to prepare nanowire scaffolds as discussed above. The resulting structures had a tunable density between 12 mg/cm$^3$, corresponding to ≈1% of the bulk value, and 4000 mg/cm$^3$ – again achievable by tuning the water suspension before freezing. A micrograph of a nanowire scaffold with a density of 80 mg/cm$^3$ is shown in Fig. 3. The resultant structure shows the random orientation of the nanowires and high porosity of the freeze-cast scaffold.

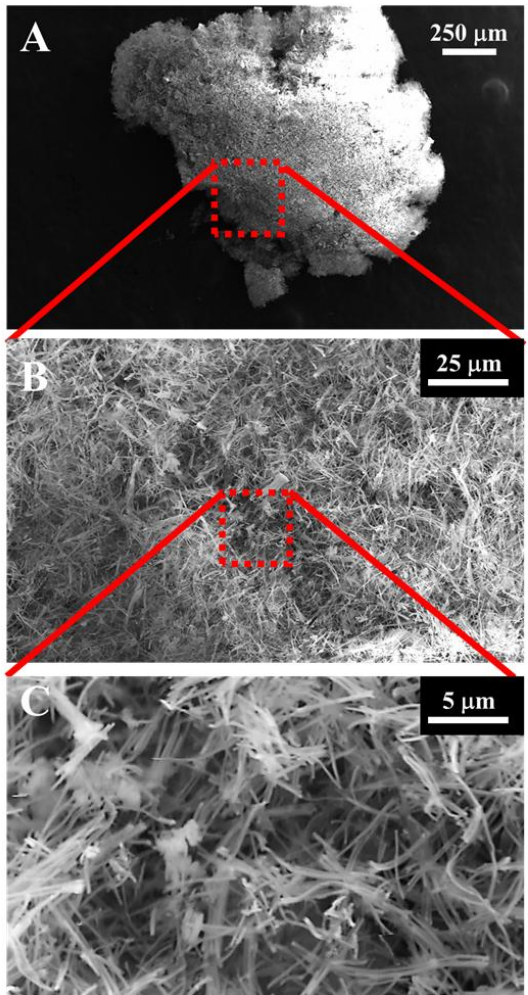

**Figure 3** SEM of the nanowire structure captured at progressively higher magnification (A-C)

*Functional Properties*

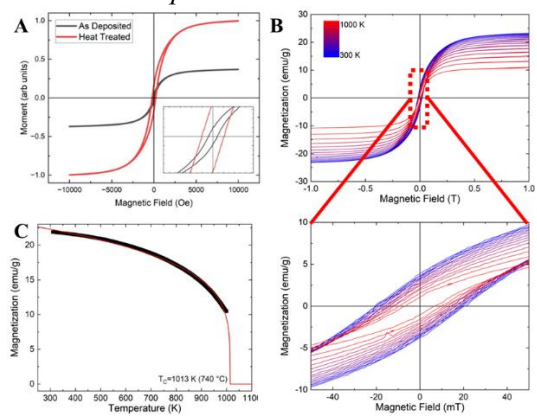

**Figure 4** (A) Major hysteresis loop for nanowires in as-deposited and heat treated conditions. Data was normalized to the post-annealed sample to allow direct comparison of the moment. (B) Hysteresis loops taken at progressively lower temperatures. (C) The MT curve with a fitted ctritcal transition curve, showing an estimated $T_C$ of 1013 K.

Magnetic measurements were performed on both as-grown and annealed the HEA nanowires. Measurements were performed by first capturing the major hysteresis loop at room temperature, then a magnetization versus temperature measurement was performed up-to 1000 K, then hysteresis loops were measured in 50 K increments as the sample was returned to room temperature. Heating to 1000 K was performed over ≈1 hr and transformed the system from as-grown, to 'Heat Treated'. The major hysteresis loops before and after heat treatment, Fig. 4a, show that both materials behave as soft ferromagnets at room temperature. The heat treatment appreciably increases the saturation magnetization and coercivity. Data is normalized to allow easy comparison between the two measurements, which were performed on the same sample; the saturation moment of the heat treated sample was 23 emu/g. The increased $M_S$ is likely a result of the non-magnetic copper being precipitated, increasing the phase fraction of the ferromagnetic elements Fe, Co, and Ni in the alloy. The heat treatment also presumably increases the ordering, which tends to increase the saturation moment in HEAs.[5] Using a dilution model the $M_S$ is expected to be ≈108 emu/g, which is higher than the experimental value, consistent with previous observations.[5] This implies the simple dilution model is not sufficient to explain our magnetic results. Indeed, Cr is an active magnetic element which can contribute to the magnetization (1) as a subtractive species aligned antiferromagnetically with the FeCoNi, (2) as a source of local frustration in the FeCoNi, and (3) as a suppressor of the ordering temperature.[32] Addressing (3) in-particular, for the binary alloys $Fe_{70}Cr_{30}$, $Co_{70}Cr_{30}$, and $Ni_{70}Cr_{30}$, only the Fe alloy is ferromagnetic at room temperature.[33-36] Magnetization versus temperature measurements were performed and show that, even at 1000 K, the sample retains magnetic polarization. Fitting the curve to a generic critical transition, the Curie temperature, $T_C$, can be estimated to be 1013 K (740 °C). For comparison, these values make the electrodeposited HEA nanowires magnetically similar to Fe ($T_C$ =770 °C), albeit with a much lower MS ($M_S^{Fe}$ = 218 emu/g).[4]

Finally, the thermal diffusivity of the HEA nanowire scaffold was measured. These measurements are particularly important as the high thermal conductivity of the scaffolded structure presents opportunities for heat exchangers which are unique among low-density materials. Conceptually, it becomes challenging to predict the propagation of heat through the scaffold structure. Using the laser lock-in thermography technique,[37, 38] the laser power (Fig. 5a) and measured temperature (Fig. 5b) are captured at a series of frequencies and fit, Fig. 5c, to generate a value for the thermal diffusivity, α, of 0.211 $mm^2 s^{-1}$ for a sample with density of 75 g/$cm^3$. Calculating the thermal conductivity from diffusivity necessitates some presumptions about heat transport which are not accurate for this system[39] and are inappropriate to compare to the steady state analog, thus will not present that value. The thermal diffusivity is comparable to bulk titanium alloys, such as Ti6242 (α=3 $mm^2 s^{-1}$), or Inconel 718 and Hastelloy C-276 (α=4 $mm^2 s^{-1}$). A diagram of the instrument and its operational setup is provided in Ref. [30] Fig. 5d.

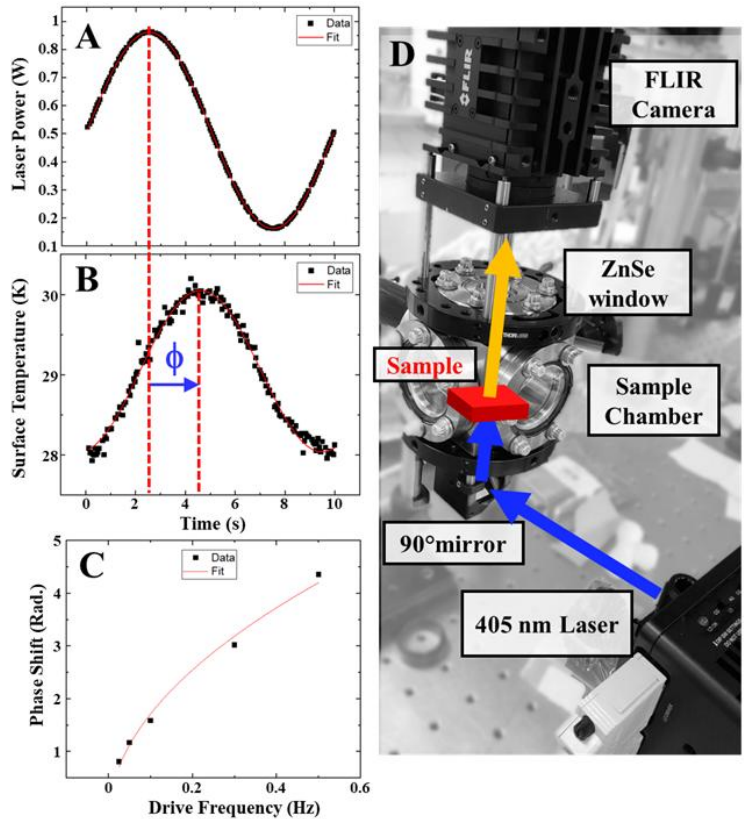

**Figure 5**: (A) frequency of the input laser. (B) the curve of the different phase shifts which extracts the thermal diffusivity. (C) frequency of the sample, which is phase shifted from the input. (D) Diagram of the thermography measurement instrument.

## Conclusion

In conclusion, we have prepared the HEA $Fe_{18}Co_{18}Ni_{27}Cr_{25}Cu_{12}$ as nanowires using electrodeposition into porous AAO membranes. The as-grown nanowires were smooth and compositionally homogeneous. After annealing, Cu precipitation was observed, forming small to medium sized grains on the surface of the nanowire. The crystal structure was revealed to be FCC, with a lattice parameter of 4.8 Å. The nanowires were freeze-cast into an ultra-low density nanowire scaffold with tunalbe densities and could be welded together, bestowing improved structural integrity, by using a reactive sintering process. The magnetic and thermal properties of the nanowire scaffolds were measured, demonstrating ferromagnetism to >1000 K, and thermal conductivity comparable to bulk aerospace alloys, at a fraction of the density. Altogether, along with the known high-temperature strength and corrosion resistance of the parent alloy, this work demonstrates a promising material for weight-critical applications, particularly in extreme environments.

## Acknowledgements

This work was supported by the Department of Energy, Office of Science, Awards DE-SC0021344 and DE-SC0026383, and the Department of Defense under award FA8650-21-2-7125. A. H. was supported by the U.S. Department of Energy, Office of Science, Basic Energy Sciences, Materials Sciences and Engineering Division. Electron microscopy was performed at the Electron Microscopy Center at the University of Tennessee, Knoxville.

TOC Figure
For Table of Contents Only

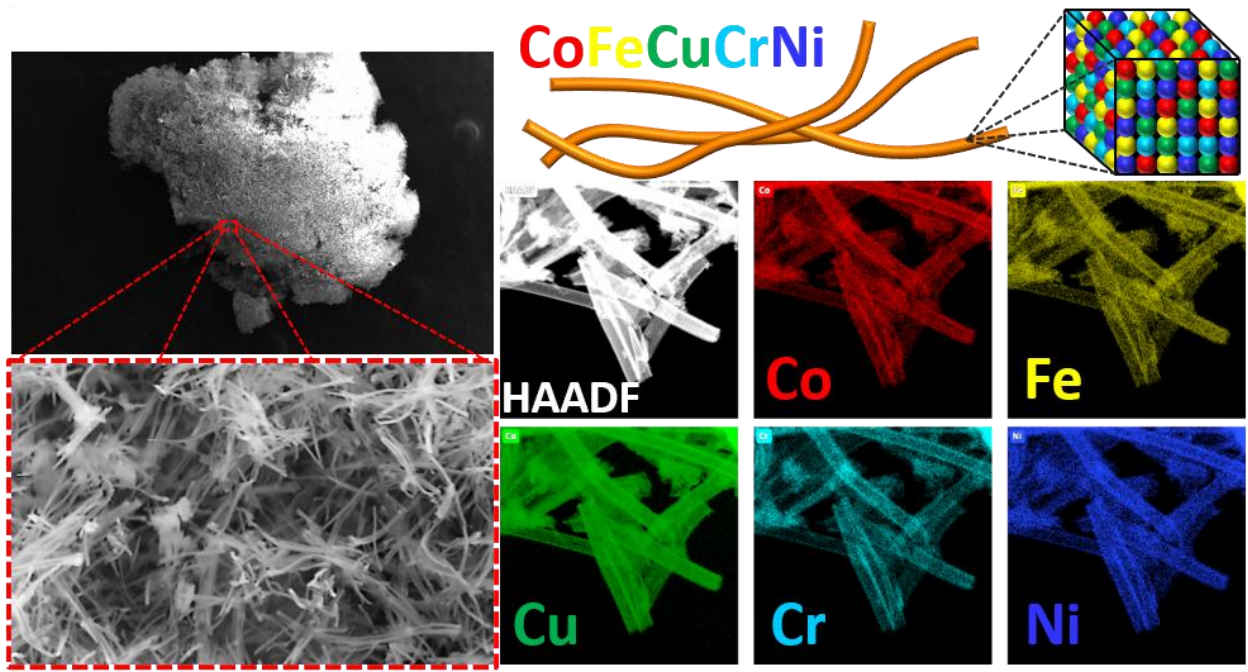